# Resolving the interfacial mechanical landscape of monolayer $NbSe_2$ grown by confined epitaxy

**Ye Wang***

Department of Applied Physics and Science Education, Eindhoven University of Technology, De Groene Loper 19, 5612 AP Eindhoven, The Netherlands

*Correspondence: y.wang19@tue.nl

**Summary**

Confined epitaxy has recently emerged as a novel methodology for controlled growth of high-quality monolayer two-dimensional (2D) materials. However, mechanistic understanding remains limited to analysing nucleus adsorption and diffusion barriers. Here, I discover that van der Waals (vdW) confinement not only affects the specific adsorption and diffusion of nuclei on confined substrates but also induces strong coupling to atomic forces at the interface. Using artificial intelligence (AI)- assisted density functional theory (DFT) calculations, I study how vdW confinement couples to the atomic motion of monolayer NbSe2 grown between c-plane sapphire and hexagonal boron nitride (h-BN). The calculated force-electronic landscape yields large mechanical forces of +0.070 and −0.04 eV $Å^{-1}$ on Nb and Se, respectively, giving rise to a unit-cell interfacial interaction energy up to 2.4 eV, comparable to reported adsorption and diffusion energies on similar substrates. The results indicate that interfacial mechanics may explain the success of confined epitaxy. The findings suggest that tuning interfacial mechanics, and related vdW mechanochemistry, could be a promising route to high-throughput, high-quality materials synthesis.

**Introduction**

Two-dimensional quantum materials provide access to superconducting, spin–valley and correlated electronic states that motivate new quantum-device architectures. Developing them beyond individual laboratory devices depends on reproducible synthesis and interface control [1]. The challenge of growing such materials is not only to increase crystal size or material purity, but also to obtain the phase and stacking configuration required for the intended function. Kinetic barriers can make these structures difficult to form or reproduce, even when a stable phase is readily accessible [1]. Understanding and controlling the atomic processes that select the growing structure is therefore important for developing these materials.

Confined epitaxy, also termed encapsulation epitaxy, offers one route to such control: the crystal grows between a substrate and a van der Waals capping layer such as graphene or hexagonal boron nitride (hBN) [2,3]. The capping layer protects the atomically thin layer but also modifies its interfacial interactions and surface chemistry [4,5]. For monolayer $NbSe_2$, which hosts charge-density-wave order and Ising superconductivity [6,7], graphene capping has protected monolayers grown by chemical vapour deposition [8], encapsulation epitaxy has enabled protected growth and integration into superconducting circuits [9], and the interfacial configuration on sapphire also influences growth [10]. The confining interfaces are therefore part of the growth environment rather than a passive protective layer.

Epitaxial growth proceeds through nucleation, incorporation at undercoordinated edges and atomic rearrangement [11]. Mechanistic analyses of these steps have mainly compared the energies of candidate configurations: substrate symmetry and directional interactions can select dichalcogenide orientations [12,13], and a preferred finite structure need not match the extended crystal [14]. These energies identify stable configurations but not how the interface affects atomic motion between them. Van der Waals interfaces do exert measurable forces, as shown by adhesion force–displacement measurements [15] and by the pressure inside trapped bubbles [16]. Within a confined crystal, these forces can differ between atomic planes and largely cancel in the net normal force, so the resultant alone does not reveal the plane-resolved response.

Such force differences matter because mechanical forces can alter chemical reactions [17]. In polymers, the response of a reactive group depends on how force is transmitted through the molecular backbone [18], and applied forces can cleave functional groups from graphene [19]. The outcome depends on atomic geometry: isotropic compression can produce anisotropic molecular strain [20], and confinement between two-dimensional crystals can induce chemical transformations in trapped molecules [21]. The same applies to a confined crystal:

forces that cancel in the resultant can still perform work when the planes they act on move relative to one another, and this work may help or hinder the rearrangements needed for incorporation. I use interfacial mechanical work to denote the work done by hBN-induced force differences along a specified displacement; a change in the energy or rate of a growth step due to this work would constitute interfacial mechanochemistry.

Here I use density functional theory (DFT) calculations to quantify these forces and their work for monolayer $NbSe_2$ grown between sapphire and hBN (Figure 1A). A human-directed, AI-assisted workflow supported the design of these comparisons, checks of the calculation records and numerical analysis (Figure 1B).

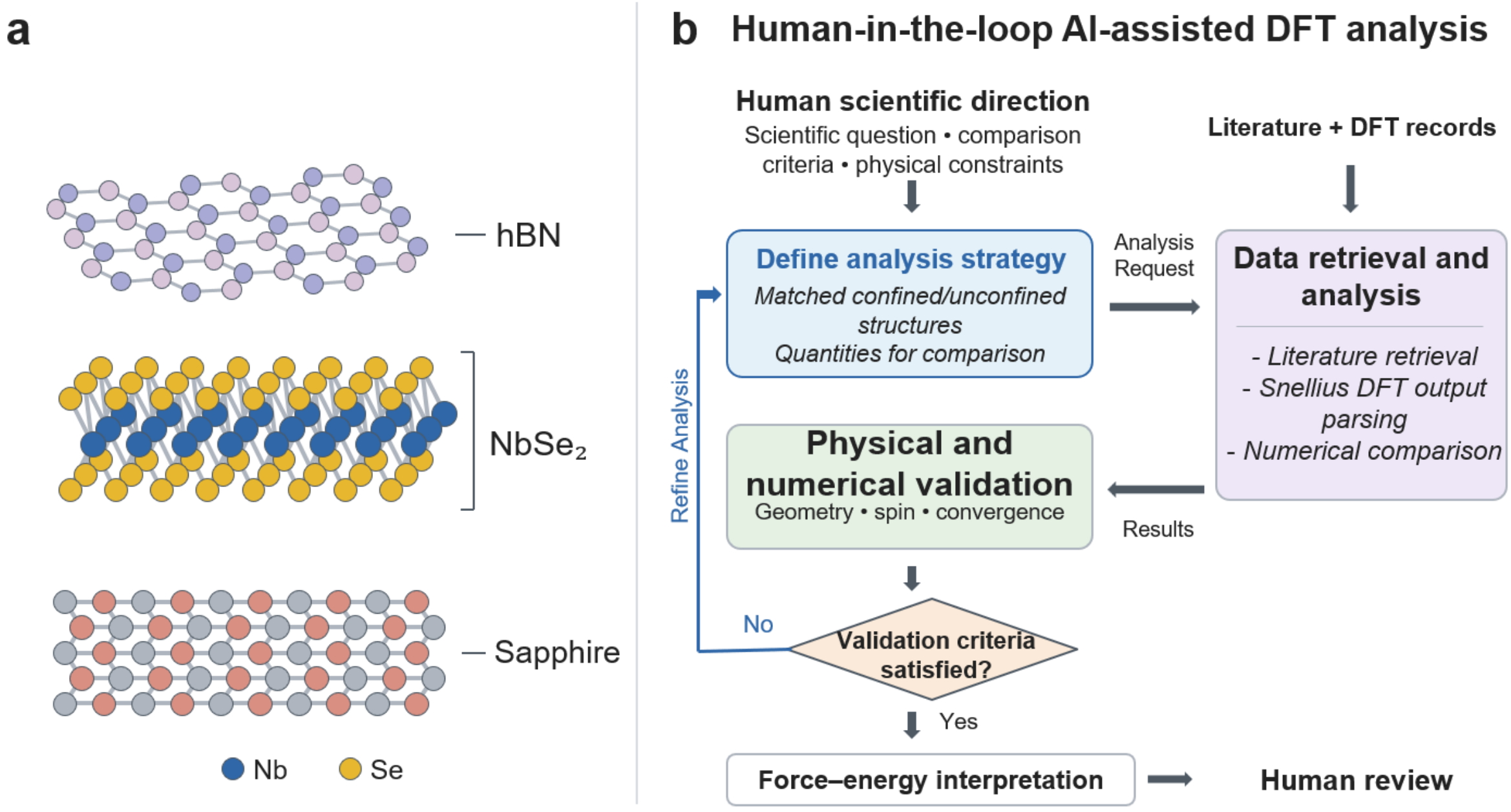

**Figure 1.** Interface structure and computational analysis. (A) Schematic of the hBN/$NbSe_2$/sapphire stack. (B) Human-directed AI-assisted workflow for matched structural comparisons, analysis and physical checks. Electronic-state selection and numerical acceptance criteria are specified in Methods.

## Results

### Sublattice-resolved interfacial forces

Figure 2 shows the additional force exerted by the hBN layer on each $NbSe_2$ atom, obtained as the difference between the confined and unconfined calculations. hBN confinement acts on the two sublattices in opposite directions. Taking positive z as pointing from sapphire towards hBN, the added normal force averages +0.070 eV $Å^{-1}$ on Nb, directed towards hBN, and −0.046 and −0.042 eV $Å^{-1}$ on the upper and lower Se planes, directed towards sapphire (Figure 2A). The maps in Figures 2C–2E show how this force varies within each plane, with color for the normal component and arrows for the in-plane component. Across all 48 $NbSe_2$ atoms, the root-mean-square force difference is 0.0567 eV $Å^{-1}$ and the largest is 0.0879 eV $Å^{-1}$.

Because these forces act in opposite directions, most of them cancel when summed over the layer. Summing the normal force over each plane gives +1.119 eV $Å^{-1}$ for Nb, −0.737 eV $Å^{-1}$ for upper Se and −0.679 eV $Å^{-1}$ for lower Se, so the net force on the whole $NbSe_2$ layer is −0.297 eV $Å^{-1}$ (Figure 2B); 88.3% of the plane forces cancel. The cancelled forces still matter, because they can do work when the Nb and Se planes move relative to each other.

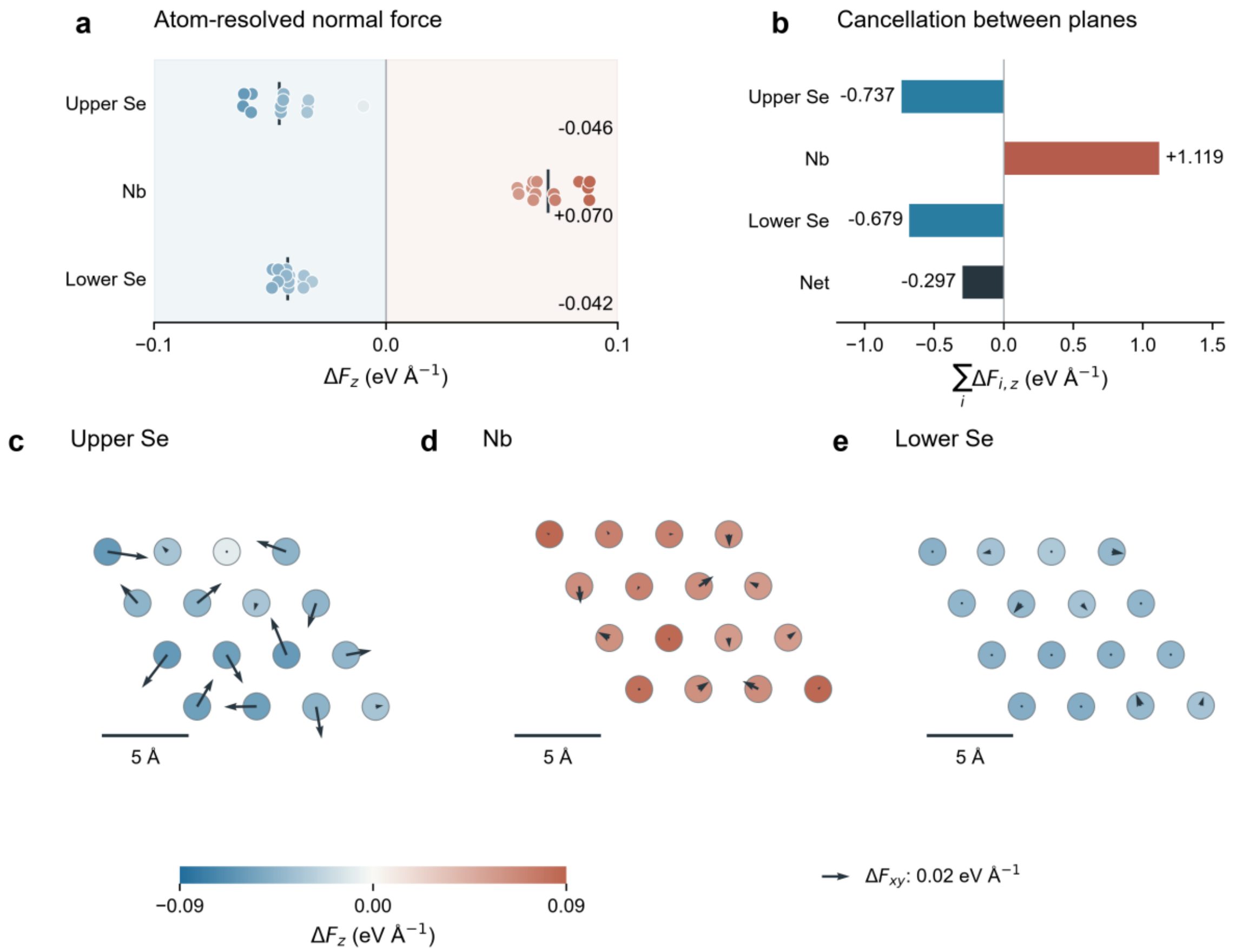

**Figure 2.** Atom-resolved interfacial force response. (A) Normal-force differences and plane means. (B) Signed plane sums and the net sum over $NbSe_2$; cancellation is 88.3%. (C–E) Upper Se, Nb and lower Se at their calculated in-plane positions, with normal force encoded by color and in-plane components by arrows. Positive z points away from sapphire. The mean normal-force changes are +0.070, −0.046 and −0.042 eV $Å^{-1}$ for Nb, upper Se and lower Se; the 48-atom RMS and maximum magnitudes are 0.0567 and 0.0879 eV $Å^{-1}$. This atom-resolved comparison uses the low-moment solutions; the high-moment force projections supporting Figure 3 are reported separately in Figure S1.

The work calculation below uses a second self-consistent solution with a larger magnetic moment (10.3–10.4 $\mu_B$ per cell), which lies about 110 meV per cell lower in energy than the solution used in Figure 2 (Methods). It shows the same opposing pattern: at the confined-derived reference geometry $G_C$, which uses the $NbSe_2$ and sapphire coordinates of the relaxed confined structure, the plane-averaged forces are +0.06312, −0.04233 and −0.03974 eV $Å^{-1}$ on Nb, upper Se and lower Se (Figure S1A). To identify which motion drives these forces, I project them onto the layer's collective displacement patterns. The projection along relative motion of Nb against Se, +0.3402 eV $Å^{-1}$, is 8 and 47 times larger in magnitude than those for rigid translation of the whole layer along z, −0.0438 eV $Å^{-1}$, or for the two Se planes moving towards or away from each other, −0.0073 eV $Å^{-1}$, respectively (Figure S1B). The hBN-induced force therefore projects predominantly onto the relative Nb–Se motion, which I use for the work calculation below. Figure S2 gives the projections onto all five patterns considered (Methods) at both $G_C$ and the unconfined-derived reference geometry $G_U$, which uses the $NbSe_2$ and sapphire coordinates of the relaxed unconfined structure.

## Force projection and interfacial energy

To turn these force differences into an energy, I define the interaction energy between the hBN layer and the underlying $NbSe_2$/sapphire structure as

$$I(R) = E_C(R) - E_U(R) - E_H,$$

where $E_C$ and $E_U$ are the total energies of the hBN-confined and unconfined structures at identical $NbSe_2$ and sapphire atomic positions R, and $E_H$ is the energy of an isolated hBN layer held in the same geometry. A negative I indicates that hBN binds to the underlying structure. For the reference structure, I = −2.39558 eV per 4 × 4 $NbSe_2$ supercell, or −0.149724 eV per $NbSe_2$ formula unit (3 × 3 × 1 k-point sampling). This quantity describes the binding of the hBN layer; its change along an internal displacement is evaluated below. The hBN-induced force on atom i ($NbSe_2$ or sapphire) is $\Delta F_i = -\partial I/\partial R_i$. To follow a specific motion, I move the atoms along a unit vector e, $R(q) = R_0 + q\,e$, and project the force difference onto that direction:

$$f_{int}(q) = \Sigma_i\, \Delta F_i(q) \cdot e_i = -dI/dq,$$

and the change in interaction energy is $\Delta I(q) = -\int_0^q f_{int}(q')\,dq'$. A positive $f_{int}$ means that the hBN-induced force drives the atoms towards positive q, so moving them towards negative q costs energy. For a fixed hBN layer and the same electronic solution, the work done by the hBN-induced force is $-\Delta I$. This work therefore quantifies, directly from the DFT total energy, how much confinement assists or resists a given motion.

I apply this to the relative Nb–Se motion identified above, starting from $G_C$. All 16 Nb atoms move along z and all 32 Se atoms move the opposite way by half as much, while sapphire and hBN stay fixed. Positive q moves Nb away from sapphire and both Se planes towards it (Figure 3A). Because the displacement vector is normalized over all 48 atoms, q is a collective coordinate: each Nb atom moves by $q/\sqrt{24} \approx 0.20q$ and each Se atom by 0.10q in the opposite direction.

At the starting geometry ($q = 0$), the projected force is +0.0766 eV $Å^{-1}$ in the confined structure and −0.2635 eV $Å^{-1}$ in the unconfined structure, so hBN contributes $f_{int}$ = +0.3402 eV $Å^{-1}$ (Figure 3B). Confinement should therefore resist motion towards negative q. I test this by moving to $q = -0.02$ Å, which shifts each Nb atom by −0.00408 Å and each Se atom by +0.00204 Å. The energy of the confined structure rises by 6.051 meV per cell, whereas that of the unconfined structure falls by 0.661 meV per cell (Figure 3C). Confinement thus imposes an energy penalty of 6.713 meV per cell, or 0.42 meV per $NbSe_2$ formula unit, for this motion.

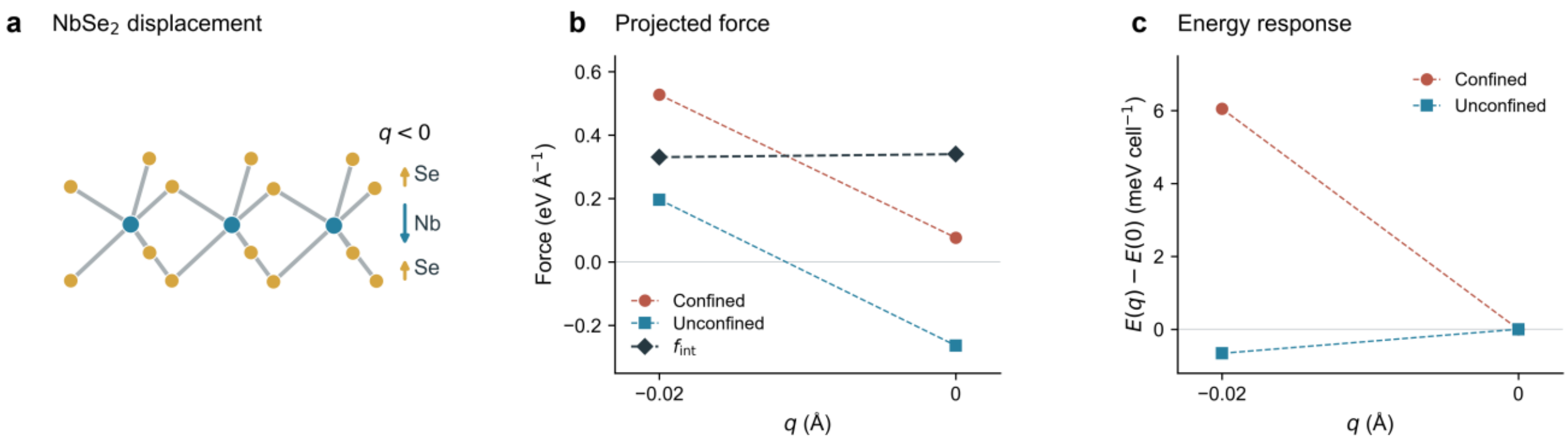

**Figure 3.** Mechanical work along an internal Nb–Se coordinate. (A) Calculated $NbSe_2$ structure with displacement arrows at $q = -0.02$ Å enlarged 250-fold; Nb blue and Se ochre. Bonds are assigned by distance. (B) Projected forces for confined and unconfined structures and their difference. (C) Energy changes relative to $q = 0$. Symbols mark the calculated points at $q = 0$ and −0.02 Å; lines are for visual guidance. The electronic solutions are the same high-moment branch used in Figures S1 and S2.

The same penalty follows from the forces alone. Integrating the force difference between the two endpoints with the trapezoidal rule gives $-W$ = +6.709 meV per cell, agreeing with the directly calculated energy change of 6.713 meV per cell to within 0.1%. The force and energy descriptions are thus mutually consistent, and the opposing Nb and Se forces carry an energy cost of 6.7 meV per cell along this coordinate even though they are largely cancelled in the net force.

### Electronic structure of the confined interface

Figure 4 compares the total density of states (DOS) of the confined and unconfined structures, using the same electronic solution as Figure 2. Both structures remain metallic, with 6.510 states $eV^{-1}$ $cell^{-1}$ at the Fermi energy in the confined structure and 6.307 in the unconfined structure (Figure 4A), whereas the DOS of isolated strained hBN vanishes at its Fermi energy, as expected for a wide-gap insulator [22] (Figure 4B). Confinement thus raises the DOS at the Fermi energy by 3.2% while preserving the metallic character of $NbSe_2$.

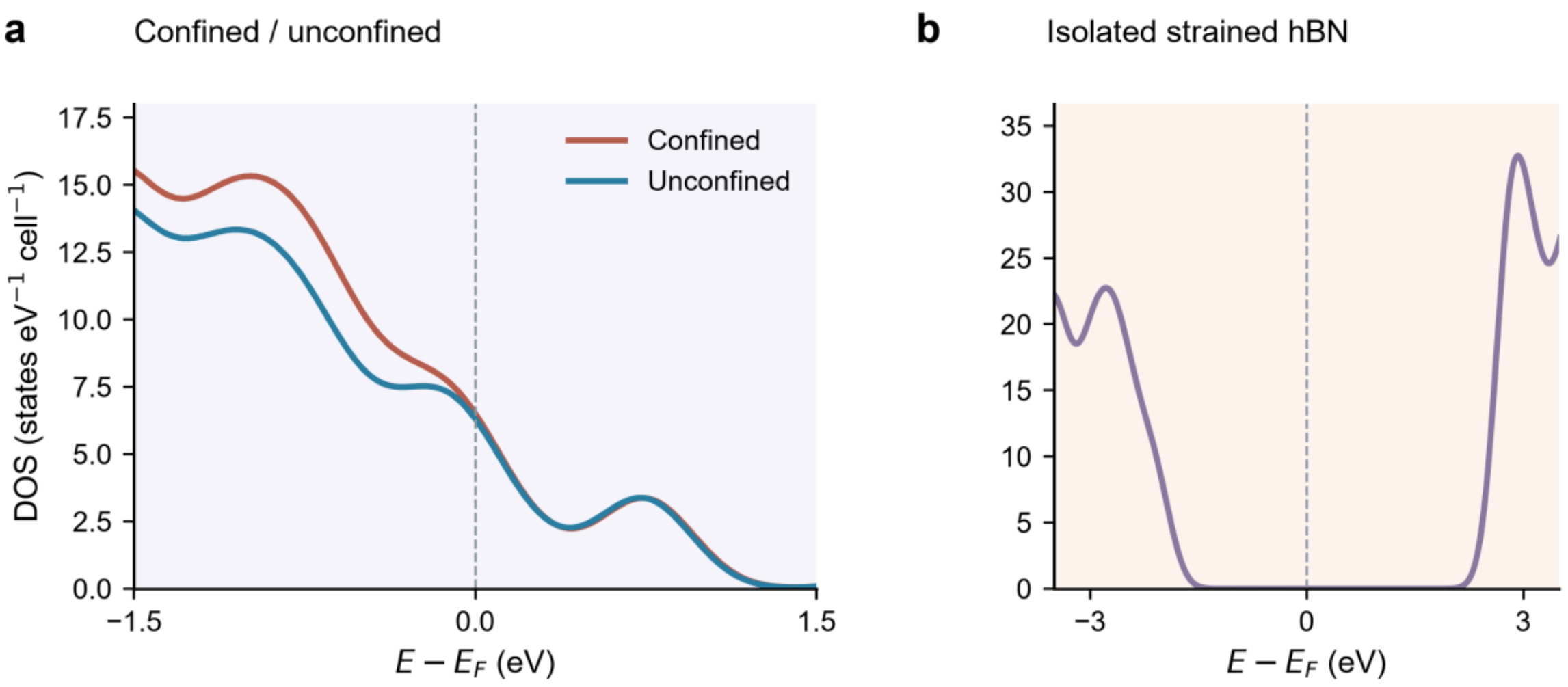

**Figure 4.** Density of states with and without hBN confinement. (A) Spin-summed total DOS of confined and unconfined structures. (B) Isolated strained hBN. Energies are referenced to each calculation's own Fermi energy; DOS is per simulation cell. Curves show the calculated samples without smoothing. Spectra correspond to the electronic solutions used in Figure 2.

## Discussion

Mechanistic descriptions of confined epitaxy have so far rested on the adsorption and diffusion energetics of the growing crystal. The present calculations show that van der Waals confinement also acts through atomic forces at the interface. hBN draws the Nb plane towards itself and pushes both Se planes towards sapphire, so that 88.3% of the plane forces cancel in the net normal force. These opposing forces are not visible in adsorption energies or in the average normal load, yet their projection onto relative Nb–Se motion (+0.340 eV $Å^{-1}$) exceeds that onto rigid translation of the layer by a factor of 8, and along this motion they perform mechanical work.

This force-based view extracts additional information from the same DFT potential-energy surface. The interaction energy of hBN with the underlying structure, −2.4 eV per supercell, measures how strongly the layer binds and is comparable to the adsorption and diffusion energies reported for encapsulation epitaxy [9]. Its negative gradient, the hBN-induced force, measures how that binding changes as the atoms move. Projected on the relative Nb–Se displacement, it gives an energy penalty of 6.7 meV per cell for shifts of 0.004 Å per Nb atom and 0.002 Å per Se atom. Binding and adsorption energies thus compare configurations, whereas interfacial forces and their work describe how the interface acts during motion between them.

The coupling between interfacial forces and internal motion offers a mechanism through which confinement could contribute to the success of confined epitaxy beyond protecting the growing layer. Growth proceeds through a sequence of atomic displacements. For each displacement, the sign of the projected interfacial force shows whether confinement assists or resists it, and its magnitude shows by how much. Evaluated along an incorporation pathway, this projection would therefore identify which growth steps confinement favors. Where this work changes the barrier or rate of growth step, the effect constitutes interfacial mechanochemistry [23,24]. Because force-dependent kinetics can involve multiple barriers [25], force-resolved analysis provides information on growth kinetics beyond that contained in adsorption energies.

The present framework establishes the force–work connection for a periodic monolayer and a well-defined internal displacement. Extending it to incorporation pathways at growing edges, including their transition states [26], will allow the contribution of interfacial work to activation barriers to be quantified directly. Electronic-state selection and numerical checks are documented in Methods and the Supplemental Information.

Interfacial mechanics therefore complements the thermodynamic description of confined epitaxy. For hBN/$NbSe_2$/sapphire, the net normal force retains 11.7% of the plane forces, yet these forces perform work along relative Nb–Se motion. Because these forces originate at the interface, they can in principle be tuned through the choice of capping layer and substrate. Interfacial forces, and the van der Waals mechanochemistry they enable, may thus provide an additional design parameter for the controlled synthesis of high-quality 2D materials.

## Methods

### Electronic-structure calculations

Calculations were performed with Quantum ESPRESSO 7.5 [27,28] using the PBE functional [29], PBE-family pseudopotentials [30,31], collinear spin polarization and Grimme D3 dispersion [32]. The wavefunction and charge-density cutoffs were 50 and 400 Ry, with a 3 × 3 × 1 k-point grid and Marzari–Vanderbilt smearing [33] of 0.020 Ry. The hexagonal supercell has an in-plane lattice parameter of 14.277 Å and a height of 45 Å including vacuum; two-dimensional electrostatic isolation was applied [34] and crystal symmetry was not imposed. Calculations were performed on the Dutch national supercomputer Snellius.

The simulation cell comprises a 3 × 3 Al-terminated α-$Al_2O_3$(0001) slab (135 atoms in nine atomic planes, 5.33 Å thick), a 4 × 4 $NbSe_2$ layer (48 atoms) and a $\sqrt{31} \times \sqrt{31}$ hBN layer (62 atoms). Pseudopotential files are listed in the Supplemental Information. With the cell and sapphire atoms fixed, $NbSe_2$ was relaxed without hBN for the unconfined structure; for the confined structure, hBN was first relaxed along z and $NbSe_2$ and hBN were then relaxed together (BFGS, Γ-point sampling, force threshold $10^{-3}$ Ry bohr$^{-1}$). The coordinates from the confined and unconfined relaxations define the confined-derived and unconfined-derived reference geometries, $G_C$ and $G_U$, which were used for the fixed-coordinate comparisons.

The in-plane supercell is fixed by the sapphire surface, and lattice mismatch is accommodated by the overlayers: principal strains of +2.2% and −1.7% in $NbSe_2$, and a biaxial strain of +2.0% in hBN relative to its calculated lattice parameter of 2.513 Å. The hBN layer is rotated by +8.95° with hollow-site registry; supercell construction details are given in the Supplemental Information.

All self-consistent calculations were converged to an estimated accuracy of $5 \times 10^{-8}$ Ry or better. The projected-force integral was evaluated with the trapezoidal rule, and its agreement with the directly calculated energy difference provides an internal consistency check. Energies are reported with Quantum ESPRESSO's finite-smearing convention.

Increasing the k-point grid from 2 × 2 × 1 to 3 × 3 × 1 changes the interaction energy by 0.04% and the force-difference RMS by 3.1%; cutoff and k-point benchmarks are given in Supplemental Tables S1 and S2.

The DOS calculations at $G_C$ used a 4 × 4 × 1 k-point grid. Spin-up and spin-down contributions were summed, each spectrum was referenced to its own Fermi energy, and the DOS at the Fermi energy was obtained by linear interpolation.

### Force and work analysis

Force differences are calculated as $F_C - F_U$ for atoms matched by index, species and position. Positive z points from sapphire towards hBN. Confined and unconfined cells contain the same 183 $NbSe_2$ and sapphire atoms, with 62 additional hBN atoms in the confined cell. Figure 2 and the DOS in Figure 4 use the low-moment solutions, and Figure 3 and Figures S1–S2 use the high-moment solutions; the two sets are analyzed separately.

The selected high-moment solutions have integrated moments of 10.33–10.41 $\mu_B$ per cell. At $G_C$ they lie below the low-moment solutions by 110.01 meV per cell in the confined structure and 110.38 meV per cell in the unconfined structure, and are therefore used for the displacement analysis. Total and absolute magnetizations were used to monitor the solutions. Figure S2 compares the same high-moment solutions at $G_U$ and $G_C$.

The normalized internal displacement has $e_{Nb,z} = 1/\sqrt{24}$ on each of 16 Nb atoms and $e_{Se,z} = -1/(2\sqrt{24})$ on each of 32 Se atoms. All other components vanish, so $\Sigma_i |e_i|^2 = 1$. The coordinate q is expressed in Å and is not mass weighted.

At each reference geometry, the force difference was projected onto five normalized $NbSe_2$ displacement patterns: opposite x and y displacements of the two Se planes with Nb fixed, rigid z translation of $NbSe_2$, opposite z displacements of the two Se planes with Nb fixed, and the internal Nb–Se pattern defined above. The energy change estimated by trapezoidal integration over these five patterns is the sum of $-[f_m(G_U) + f_m(G_C)]\Delta q_m/2$, where $f_m$ is the projected force difference along pattern m and $\Delta q_m$ is the change in that coordinate from $G_U$ to $G_C$; the difference from the directly calculated interaction-energy change is reported.

Plane-averaged normal-force changes were recovered from the three normalized projections along z. If $S_{Nb}$, $S_u$ and $S_l$ denote sums over the 16 Nb, upper-Se and lower-Se atoms, respectively, the relative Nb–Se, rigid-z and Se-breathing projections obey $\sqrt{24}\,P_b = S_{Nb} - (S_u + S_l)/2$, $\sqrt{48}\,P_z = S_{Nb} + S_u + S_l$ and $\sqrt{32}\,P_s = S_u - S_l$. These three linear equations uniquely determine the plane sums; division by 16 gives the means.

The quantitative figures, including the plane means reconstructed from normalized force projections, were prepared from the calculated atomic coordinates, forces, energies and DOS. OpenAI Codex (recorded model identifiers gpt-5.6-sol, gpt-6-astra and gpt-6-sol; August–September 2026) assisted with literature retrieval, comparison design, record checks, analysis code, figures and drafting; I set the scientific question and comparison conditions and checked numerical results against the electronic-structure records.

**Resource availability**

Requests for further information should be directed to the lead contact, Ye Wang (y.wang19@tue.nl). Materials availability: this study did not generate new materials. Data and code availability: the data reported in this paper and the custom Python scripts used for force-difference analysis, displacement-mode projection, endpoint-work evaluation, DOS processing and figure preparation will be shared by the lead contact upon request. Quantum ESPRESSO was used for the electronic-structure calculations; the custom scripts perform analysis and visualization of its outputs. Any additional information required to reanalyze the data reported in this paper is available from the lead contact upon request.

## Acknowledgments

This work was supported by the Technische Universiteit Eindhoven startup grant RF204648 and the Dutch Research Council (NWO), grant no. EINF-19934.

## References

1. Liebregts, M. & Wang, Y. Two-Dimensional Materials toward CMOS-Compatible Scalable Quantum Hardware. arXiv:2609.16172 (2026).

2. Al Balushi, Z. Y. et al. Two-dimensional gallium nitride realized via graphene encapsulation. Nature Materials 15, 1166–1171 (2016). https://doi.org/10.1038/nmat4742

3. Briggs, N. et al. Atomically thin half-van der Waals metals enabled by confinement heteroepitaxy. Nature Materials 19, 637–643 (2020). https://doi.org/10.1038/s41563-020-0631-x

4. Geim, A. K. & Grigorieva, I. V. Van der Waals heterostructures. Nature 499, 419–425 (2013). https://doi.org/10.1038/nature12385

5. Fu, Q. & Bao, X. Surface chemistry and catalysis confined under two-dimensional materials. Chemical Society Reviews 46, 1842–1874 (2017). https://doi.org/10.1039/C6CS00424E

6. Xi, X. et al. Strongly enhanced charge-density-wave order in monolayer $NbSe_2$. Nature Nanotechnology 10, 765–769 (2015). https://doi.org/10.1038/nnano.2015.143

7. Xi, X. et al. Ising pairing in superconducting $NbSe_2$ atomic layers. Nature Physics 12, 139–143 (2016). https://doi.org/10.1038/nphys3538

8. Wang, H. et al. High-quality monolayer superconductor $NbSe_2$ grown by chemical vapour deposition. Nature Communications 8, 394 (2017). https://doi.org/10.1038/s41467-017-00427-5

9. Zheng, X. et al. Encapsulation epitaxy of air-stable 2D superconductors for quantum circuits. Nature 656, 349–356 (2026). https://doi.org/10.1038/s41586-026-10865-1

10. Yin, S. et al. Impact of interfacial configurations on the growth of few-layer $NbSe_2$ on sapphire. Nano Letters 25, 10603–10610 (2025). https://doi.org/10.1021/acs.nanolett.5c02390

11. Dong, J., Zhang, L., Dai, X. & Ding, F. The epitaxy of 2D materials growth. Nature Communications 11, 5862 (2020). https://doi.org/10.1038/s41467-020-19752-3

12. Zheng, P. et al. Universal epitaxy of non-centrosymmetric two-dimensional single-crystal metal dichalcogenides. Nature Communications 14, 592 (2023). https://doi.org/10.1038/s41467-023-36286-6

13. Li, T. et al. Epitaxial growth of wafer-scale molybdenum disulfide semiconductor single crystals on sapphire. Nature Nanotechnology 16, 1201–1207 (2021). https://doi.org/10.1038/s41565-021-00963-8

14. Zong, C. et al. van der Waals magnetic single-atomic chains kinetically accessible on surfaces. Newton, 100681 (2026). https://doi.org/10.1016/j.newton.2026.100681

15. Rokni, H. & Lu, W. Direct measurements of interfacial adhesion in 2D materials and van der Waals heterostructures in ambient air. Nature Communications 11, 5607 (2020). https://doi.org/10.1038/s41467-020-19411-7

16. Khestanova, E. et al. Universal shape and pressure inside bubbles appearing in van der Waals heterostructures. Nature Communications 7, 12587 (2016). https://doi.org/10.1038/ncomms12587

17. Davis, D. A. et al. Force-induced activation of covalent bonds in mechanoresponsive polymeric materials. Nature 459, 68–72 (2009). https://doi.org/10.1038/nature07970

18. Klukovich, H. M. et al. A backbone lever-arm effect enhances polymer mechanochemistry. Nature Chemistry 5, 110–114 (2013). https://doi.org/10.1038/nchem.1540

19. Felts, J. R. et al. Direct mechanochemical cleavage of functional groups from graphene. Nature Communications 6, 6467 (2015). https://doi.org/10.1038/ncomms7467

20. Yan, H. et al. Sterically controlled mechanochemistry under hydrostatic pressure. Nature 554, 505–510 (2018). https://doi.org/10.1038/nature25765

21. Vasu, K. S. et al. Van der Waals pressure and its effect on trapped interlayer molecules. Nature Communications 7, 12168 (2016). https://doi.org/10.1038/ncomms12168

22. Cassabois, G., Valvin, P. & Gil, B. Hexagonal boron nitride is an indirect bandgap semiconductor. Nature Photonics 10, 262–266 (2016). https://doi.org/10.1038/nphoton.2015.277

23. Bell, G. I. Models for the specific adhesion of cells to cells. Science 200, 618–627 (1978). https://doi.org/10.1126/science.347575

24. Ribas-Arino, J. & Marx, D. Covalent mechanochemistry: theoretical concepts and computational tools with applications to molecular nanomechanics. Chemical Reviews 112, 5412–5487 (2012). https://doi.org/10.1021/cr200399q

25. Tian, Y. et al. Model studies of force-dependent kinetics of multi-barrier reactions. Nature Communications 4, 2538 (2013). https://doi.org/10.1038/ncomms3538

26. Henkelman, G., Uberuaga, B. P. & Jónsson, H. A climbing image nudged elastic band method for finding saddle points and minimum energy paths. Journal of Chemical Physics 113, 9901–9904 (2000). https://doi.org/10.1063/1.1329672

27. Giannozzi, P. et al. QUANTUM ESPRESSO: a modular and open-source software project for quantum simulations of materials. Journal of Physics: Condensed Matter 21, 395502 (2009). https://doi.org/10.1088/0953-8984/21/39/395502

28. Giannozzi, P. et al. Advanced capabilities for materials modelling with Quantum ESPRESSO. Journal of Physics: Condensed Matter 29, 465901 (2017). https://doi.org/10.1088/1361-648X/aa8f79

29. Perdew, J. P., Burke, K. & Ernzerhof, M. Generalized gradient approximation made simple. Physical Review Letters 77, 3865–3868 (1996). https://doi.org/10.1103/PhysRevLett.77.3865

30. Dal Corso, A. Pseudopotentials periodic table: from H to Pu. Computational Materials Science 95, 337–350 (2014). https://doi.org/10.1016/j.commatsci.2014.07.043

31. Garrity, K. F., Bennett, J. W., Rabe, K. M. & Vanderbilt, D. Pseudopotentials for high-throughput DFT calculations. Computational Materials Science 81, 446–452 (2014). https://doi.org/10.1016/j.commatsci.2013.08.053

32. Grimme, S., Antony, J., Ehrlich, S. & Krieg, H. A consistent and accurate ab initio parametrization of density functional dispersion correction (DFT-D) for the 94 elements H–Pu. Journal of Chemical Physics 132, 154104 (2010). https://doi.org/10.1063/1.3382344

33. Marzari, N., Vanderbilt, D., De Vita, A. & Payne, M. C. Thermal contraction and disordering of the Al(110) surface. Physical Review Letters 82, 3296–3299 (1999). https://doi.org/10.1103/PhysRevLett.82.3296

34. Sohier, T., Calandra, M. & Mauri, F. Density functional perturbation theory for gated two-dimensional heterostructures: theoretical developments and application to flexural phonons in graphene. Physical Review B 96, 075448 (2017). https://doi.org/10.1103/PhysRevB.96.075448